\documentclass[prc,aps,showpacs]{revtex4}
\begin{document}
 
\newcommand{\be}{\begin{eqnarray}}
\newcommand{\ee}{\end{eqnarray}}

\title{Relativistic corrections to the nuclear Schiff moment}

\author{ V.F.Dmitriev$^{1,2}$,and V.V. Flambaum$^2$  }

\affiliation{$^1$
Budker Institute of Nuclear Physics, Novosibirsk, Russia
}

\affiliation{$^2$
 School of Physics, The University of New South Wales, Sydney NSW
2052,
Australia
}

\date{\today}

%*****************************************************************
\begin{abstract}

Parity and time invariance violating ($P,T$-odd)
atomic electric dipole moments (EDM) are induced by interaction between
atomic electrons and nuclear $P,T$-odd moments which are produced
 by $P,T$-odd nuclear forces. The nuclear EDM is screened by atomic
 electrons. The EDM of a non-relativistic atom with closed electron
 subshells is induced by the nuclear Schiff moment.
For heavy relativistic atoms EDM is induced by the nuclear local
 dipole moments which differ by 10-50\% from the Schiff moments
 calculated previously. 
 We calculate the local dipole moments  for ${^{199}{\rm Hg}}$
and ${^{205}{\rm Tl}}$ where the most accurate atomic
\cite{fortson} and molecular \cite{hinds}
 EDM measurements have been performed.
\end{abstract} 
\vspace{1cm}

\maketitle
\vspace{0.1in}
\section{Introduction}
The best limits on Parity(P) and Time(T) invariance violating nuclear
 forces are obtained from the measurement of  electric dipole
moment (EDM) of 
${^{199}{\rm Hg}}$ atom 
\cite{fortson}.
The mechanism of this EDM generation is the following. 
$P,T$-odd nuclear forces produce $P,T$-odd nuclear moments.
 In turn, these moments can induce electric dipole moments 
 in atoms through the mixing of electron wave functions of opposite
parity. The EDM of a point-like nucleus according to the Schiff theorem 
\cite{pr50,sch63,san67} is 
completely screened by atomic electrons. However, due to finite nuclear size,
some dipole component of the electrostatic potential still exists. It is 
generated by the Schiff moment \cite{skf84}, the higher moment in the 
dipole density distribution. 

The electrostatic potential produced by the Schiff moment is usually 
presented in the form \cite{skf84}
\be \label{1}
\phi({\bf R}) = 4\pi{\bf S}\cdot \mbox{\boldmath $\nabla$}\delta({\bf R}),
\ee
where $\bf S$ is the Schiff moment. Presence of the gradient of $\delta
({\bf R})$ means that the matrix element of the potential given by Eq.
(\ref{1}) between $s$- and $p$-atomic states is proportional to 
$({\mathbf \nabla}\psi^\dagger_p\psi_s)_{R=0}$. However, at high $Z$ the
relativistic wave functions of atomic electrons vary significantly
 ($\sim Z^2\alpha^2$; for ${^{199}{\rm Hg}}$  $Z^2\alpha^2=0.34$)
inside the nucleus. This produce a corrections from higher moments of the 
dipole density distribution leading to a {\it local dipole moment} \cite{fg02}
$$
{\bf L} =e\sum_{k=1}^\infty\frac{b_k}{b_1}\frac{1}{(k+1)(k+4)}\left[\langle
{\bf r}r^{k+1}\rangle-\frac{k+4}{3}\langle{\bf r}\rangle\langle r^{k+1}\rangle
\right] 
$$
\be \label{2}
= L {\bf I}/I.
\ee
The summation is carried  over odd powers of $k$,   $k=1,3,5,...$
The first term ($k=1$) in Eq.(\ref{2}) is just the Schiff moment and the first nonzero
correction ($k=3$) to it comes from $b_3/b_1$. The factors $b_3/b_1$ being both of the
order of $Z^2\alpha^2$  differ for
$p_{1/2}$ and $p_{3/2}$ atomic states \cite{fg02}
\be \label{3}
\frac{b_3}{b_1}= -\frac{3}{5}\frac{Z^2\alpha^2}{R_N^2},
\ee 
for $s-p_{1/2}$ matrix element, and 
\be \label{4}
\frac{b_3}{b_1}= -\frac{9}{20}\frac{Z^2\alpha^2}{R_N^2},
\ee 
for $s-p_{3/2}$ matrix element. Here $R_N$ is the charge radius of uniformly 
charged nucleus. 

    When doing atomic EDM calculations one should use a corrected
P,T-odd interaction Hamiltonian $W_{PT}$ for relativistic atomic electrons
containing the  local dipole moment \cite{fg02}:
\be \label{l1}
W_{PT}/(-e) \equiv \phi_L({\bf R}) = 4\pi{\bf L}\cdot \mbox{\boldmath $\nabla$}\delta({\bf R}),
\ee
To use this formula one must assume that the local dipole moment
is placed in the centre of a finite-size nucleus.

The aim of the present work is to calculate the local dipole moment
$L$ for two most important nuclei ${^{199}{\rm Hg}}$ (with valence
neutron) and ${^{205}{\rm Tl}}$ (with valence proton).

\section{Outlines of the theory}
In calculations we used a finite range P- and T-violating nucleon-nucleon 
interaction of the form
$$
W({\bf r}_a - {\bf r}_b) = -\frac{g_s}{8\pi m_p}\left[ (g_0\mbox{\boldmath
 $\tau$}_a\cdot\mbox{\boldmath $\tau$}_b   +
g_2(\mbox{\boldmath $\tau$}_a\cdot \mbox{\boldmath $\tau$}_b
 -3\tau_a^z\tau_b^z))   \right.
$$
\begin{equation} \label{5}
\left.( \mbox{\boldmath $\sigma$}_a -
\mbox{\boldmath $\sigma$}_b) + g_1(\tau_a^z\mbox{\boldmath $\sigma$}_a  - 
\tau_b^z\mbox{\boldmath
$\sigma$}_b)\right]  \cdot \mbox{\boldmath $\nabla$}_a \frac{e^{-m_{\pi}
r_{ab}}}{r_{ab}},
\end{equation}
where $m_p$ is the proton mass and $r_{ab}=|{\bf r}_a-{\bf r}_b|$.

For a nucleus having an odd proton there are two kinds of
contributions to the local dipole moment even in the absence of a
strong residual interaction between the odd proton and the nucleons in
the core. One contribution comes from the odd proton.
Another contribution comes from the core nucleons.
We start from the odd proton contribution. The weak interaction, 
Eq.(\ref{5}), generates a weak correction to the nuclear
mean field which, in turn, produces a correction 
$\delta\psi_\nu({\bf
r})$ to the single particle wave function $\psi_\nu({\bf r})$ of the
odd proton. With this correction the expectation value of the local dipole
moment is
\begin{equation} \label{6}
L_{sp}
= \langle\delta\psi_\nu|\hat{L}_z|\psi_\nu\rangle
+\langle\psi_\nu|\hat{L}_z|\delta \psi_\nu\rangle.
\end{equation}
Note that this single particle contribution is absent for neutron odd
nuclei.

 Another
type of contribution to the local dipole moment comes from off diagonal
matrix elements of the two body weak interaction.
\begin{equation} \label{7}
L_{core}=\sum_{\nu_1\nu_2}\langle\nu,\nu_1|W|\nu_2,\nu\rangle
\frac{n_{\nu_1} - n_{\nu_2}}{\epsilon_{\nu_1} -
\epsilon_{\nu_2}}\langle\nu_2|\hat{L}_z |\nu_1\rangle,
\end{equation}
where $n_\nu$ are the occupation numbers and $\epsilon_\nu$ are the
single particle energies. Here the operator $\hat{L}_z$ acts on the core
protons. The magnitudes of $S_{sp}$ and $S_{core}$ are
comparable for proton odd nuclei. This is in contrast to the P odd and T
even  nuclear anapole moment where the core contribution is
 smaller by a factor $1/A^{1/3}$.

When the strong residual interaction leading to core polarization is taken into
account the contributions to
the local dipole moment can be written as a sum of three terms \cite{ds1,dsa}
\begin{equation} \label{8}
L = \langle\delta\psi_\nu|\tilde{L}_z|\psi_\nu\rangle +
\langle\psi_\nu|\tilde{L}_z|\delta\psi_\nu\rangle +\langle\nu|\delta L|\nu\rangle.
\end{equation}
The first two terms are those where the weak correction enters via the
odd particle wave function. $\tilde{L}$ satisfies the equation
\begin{equation} \label{9}
\langle\nu'|\tilde{\bf L}|\nu\rangle =\langle\nu'|{\bf L}|\nu\rangle+\sum_{\nu_1\nu_2}
\langle\nu'\nu_1|F|\nu_2\nu\rangle
\frac{n_{\nu_1} - n_{\nu_2}}{\epsilon_{\nu_1} -
\epsilon_{\nu_2}}\langle\nu_2|\tilde{\bf L}|\nu_1\rangle.
\end{equation}
 Here only
the residual strong interaction $F$ enters in the equation. The effects of
the weak interaction are entirely in the wave functions of the odd
proton. Eq.(\ref{9}) describes the well known effect of renormalization
of nuclear moments due to coupling with particle-hole states in the
core. The third term in (\ref{8}) satisfies the equation
\begin{equation} \label{10}
\langle\nu'|\delta {\bf L}|\nu\rangle =\langle\nu'|\delta {\bf
L}_0|\nu\rangle+\sum_{\nu_1\nu_2}
\langle\nu'\nu_1|F|\nu_2\nu\rangle
\frac{n_{\nu_1} - n_{\nu_2}}{\epsilon_{\nu_1} -
\epsilon_{\nu_2}}\langle\nu_2|\delta{\bf L}|\nu_1\rangle,
\end{equation}
that looks similar to (\ref{9}). However, the inhomogenious term
$\delta {\bf S}_0$ is completely different, namely
$$
\langle\nu'|\delta{\bf L}_0|\nu\rangle=\sum_{\nu_1\nu_2}
\langle\nu'\nu_1|W|\nu_2\nu\rangle
\frac{n_{\nu_1} - n_{\nu_2}}{\epsilon_{\nu_1} -
\epsilon_{\nu_2}}\langle\nu_2|\tilde{\bf L}|\nu_1\rangle  +
$$
$$
\sum_{\nu_1\nu_2}
(\langle\nu'\delta\psi_{\nu_1}|F|\nu_2\nu\rangle+
\langle\nu'\nu_1|F|\delta\psi_{\nu_2}\nu\rangle)
\frac{n_{\nu_1} - n_{\nu_2}}{\epsilon_{\nu_1} -
\epsilon_{\nu_2}}\langle\nu_2|\tilde{\bf L}|\nu_1\rangle +
$$
\begin{equation} \label{11}
\sum_{\nu_1\nu_2}
\langle\nu'\nu_1|F|\nu_2\nu\rangle
\frac{n_{\nu_1} - n_{\nu_2}}{\epsilon_{\nu_1} -
\epsilon_{\nu_2}}(\langle\delta\psi_{\nu_2}|\tilde{\bf L}|\nu_1\rangle
+ \langle\nu_2|\tilde{\bf L}|\delta\psi_{\nu_1}\rangle).
\end{equation}
The first term on the rhs of Eq.(\ref{11}) is the core contribution
(\ref{7}), where instead of the bare local dipole moment operator (\ref{2})
enters the renormalized operator $\tilde{\bf L}$. The second and the
third terms correspond to additional contributions where the weak
interaction enters via corrections to the intermediate single particle
states $|\nu_1\rangle$ and $|\nu_2\rangle$. Equations (9-11) describe all
 of the core polarization effects.

Results of the calculations of the Schiff moment and the first relativistic 
correction for the proton odd nucleus $^{205}$Tl and the neutron odd nucleus 
$^{199}$Hg are shown in Table I.
  We see that usually the difference between the local dipole moment (which
is actually measured in atomic EDM experiments) and the Schiff moment
is not very large, about 10-20\%. However, the correction $L'$ may reduce
 the result two times if the value of the Schiff moment is anomalously small
(see the $g_2$ contribution for $^{205}$Tl).
\begin{table}
\begin{tabular}{c|c|c|c|c|c|c|c|c}
\multicolumn{5}{c}{$^{205}$Tl}  & \multicolumn{4}{|c}{$^{199}$Hg} \\ 
\hline 
 & $S_0$ & $L'_0/S_0$& $S_{tot}$&$L'_{tot}/S_{tot}$&$S_0$ & $L'_0/S_0$& $S_{tot}$&$L'_{tot}/S_{tot}$\\
\hline
$g_0$ & -0.075 &-0.09 &0.014 &-0.18&-0.085&-0.1&-0.006&-0.05\\
\hline
$g_1$&-0.028&-0.39&-0.082&-0.18&-0.085&-0.1&-0.036&-0.15\\
\hline
$g_2$&0.237&-0.08&-0.007&-0.51&0.17&-0.1&0.019&-0.08\\
\hline
\end{tabular}
\caption{Schiff moment $S$ and the ratio of relativistic correction $L'$ to
$S$ ($L=S+L'$) for proton odd
$^{205}$Tl and neutron odd $^{199}$Hg nuclei. $S_0$ and $L'_0$ are the bare 
values, without strong residual nuclear forces. $S_{tot}$ and $L'_{tot}$ are
the total results with full account of core polarization effects. The values
of $L'/S$ are given for $s-p_{1/2}$ transition.
 For $s-p_{3/2}$ transition they differ by the factor 3/4.
To obtain values of $L$ and $S$ one should sum up contributions
of three interaction constants $g_0$, $g_1$ and $g_2$.}
\end{table}

\acknowledgments

This work was supported by the Australian Research Council
and G. Godfrey fund.

\end{document}